\documentclass[prd,superscriptaddress,floatfix,amsfonts,amssymb,amsmath,showpacs,twocolumn]{revtex4-1}
\usepackage{bm}
\usepackage{amsfonts}
\usepackage{latexsym}
\usepackage[latin1]{inputenc}
\usepackage{graphicx}
\usepackage{amsmath}
\usepackage{palatino}
\usepackage{mathpazo}
\usepackage{textcomp}
\usepackage{booktabs}
\usepackage{dcolumn}
\usepackage{hyperref}
\hypersetup{colorlinks,citecolor=blue}
\usepackage{amsmath}
\usepackage{xcolor}
\usepackage{mathtools}
\usepackage{multirow}
\usepackage{float}
\usepackage{orcidlink}

\allowdisplaybreaks[1]

\begin{document}
\color{red}

\title{Constraining $f(Q,T)$ gravity from energy conditions}

\author{Simran Arora\orcidlink{0000-0003-0326-8945}}
 \email{dawrasimran27@gmail.com}
\affiliation{ Department of Mathematics, Birla Institute of Technology and Science-Pilani,\\ Hyderabad Campus,Hyderabad-500078, India}
\author{J.R.L. Santos\orcidlink{0000-0002-9688-938X}}
\email{joaorafael@df.ufcg.edu.br}
\affiliation{UFCG - Universidade Federal de Campina Grande - Unidade Acad\^{e}mica de F\'isica,  58429-900 Campina Grande, PB, Brazil.}
\author{P.K. Sahoo\orcidlink{0000-0003-2130-8832}}
 \email{pksahoo@hyderabad.bits-pilani.ac.in}
\affiliation{ Department of Mathematics, Birla Institute of Technology and Science-Pilani,\\ Hyderabad Campus,Hyderabad-500078, India}
\begin{abstract}
We are living in a golden age for experimental cosmology. New experiments with high accuracy precision are been used to constrain proposals of several theories of gravity, as it has been never done before. However, important roles to constrain new theories of gravity in a theoretical perspective are the energy conditions. Throughout this work, we carefully constrained some free parameters of two different families of $f(Q,T)$ gravity using different energy conditions.  This theory of gravity combines the gravitation effects through the non-metricity scalar function $Q$, and manifestations from the quantum era of the Universe in the classical theory (due to the presence of the trace of the energy-momentum tensor  $T$). Our investigation unveils the viability of $f(Q,T)$ gravity to describe the accelerated expansion our Universe passes through. Besides, one of our models naturally provides a phantom regime for dark energy and satisfies the dominant energy condition. The results here derived strength the viability of $f(Q,T)$ as a promising complete theory of gravity, lighting a new path towards the description of the dark sector of the Universe. 

\end{abstract}

\maketitle

\section{Introduction}\label{sec1}

Since the remarkable measurements from Supernova Cosmology Project \cite{riess_98} and High Redshift Supernova Team \cite{perl_99}, we have consciously known that our Universe passes through an accelerated phase of expansion, whose agent is named dark energy. Dark energy corresponds to approximately $70\,\%$ of the content of the so-called dark sector of the Universe, and its understanding is one of the actual biggest problems in science. A simple path to describe the nature of the dark energy consists in to add a cosmological constant to Einstein's General Relativity (GR), yielding to the $\Lambda$CDM model. However, the cosmological constant brings several other issues related to its nature. Among them, we highlight the cosmic coincidence problem, and its huge discrepancy between cosmological observations and quantum field theory predictions, which is about $120$ orders of magnitude \cite{adler_95}.

Apart from these listed problems, GR stills the most well succeed theory to describe the Universe. It was confirmed by several surveys such as PLANCK Collaboration \cite{Planck/2018}, Dark Energy Survey \cite{Des/2018}, besides the recent beautiful measurements of gravitational waves from LIGO/VIRGO Collaboration \cite{lv_papers}, and the first image of a black hole obtained by the Event Horizon Telescope \cite{eht_papers}. Although, GR does not yield to a renormalizable quantum theory for gravity, opening space to several alternative theories desiring to describe gravity at a quantum level. 

A promising theory of gravity was introduced by Jimenez et al. \cite{Jimenez/2018}, and called symmetric teleparallel gravity or $f(Q)$, where the gravitation interaction is mediated by the non-metricity term $Q$. Such a theory rapidly inspired several works and it has been constantly tested. Among such tests, we highlight the work done by Lazkoz et al \cite{Lazkoz/2019}, were several $f(Q)$ models were constrained through redshift comparison with data from the expansion rate, Type Ia Supernovae, Quasars, Gamma-Ray  Bursts, Baryon  Acoustic  Oscillations data,  and  Cosmic  Microwave  Background distance. Besides, the $f(Q)$ model also unveiled a compatible description of an accelerated phase when submitted to energy conditions constraints as shown in \cite{Mandal/2020}.

As Jimenez noted in his work, $f(Q)$ theories share the background equations with $f(T)$ theories, which at the perturbative level often lead to strong coupling problems. It is shown that in $f(Q)$ models on a general FLRW background, the strong coupling issues experienced in $f(T)$ theories are absent. However, on maximally symmetric backgrounds such as Minkowski and de-Sitter, they do appear. It can be ensured that the same coupling issues can also be found in $f(Q, T)$ on a maximally symmetric background.  Similar to $f(Q)$, two scalar degrees of freedom that are absent in $f(T)$ are also propagated on large scales by the $f(Q, T)$ model. These two degrees of freedom are the ones that vanish around maximally symmetric backgrounds, and thus cause the problem of strong-coupling. Second, a gauge symmetry given by a restricted diffeomorphism is retained by the maximally symmetric backgrounds. As a result of a residual gauge symmetry that then roots the strong coupling issues, these findings allow us to provide a better understanding of the disappearance of degrees of freedom around these backgrounds.
 
In the search for a complete theory of gravity emerged the $f(Q,T)$, recently presented by Yixin Xu et al. \cite{Yixin/2019}. Such a theory couples the gravitation effects through the non-metricity function $Q$, and manifestations from the quantum domain in the classical theory (due to the presence of the trace of the energy-momentum tensor  $T$). This new formulated $f(Q, T)$ gravity is constructed in a manner similar to $f(R, T)$\cite{Harko_Lobo/2018, Harko/2011}, but with the standard Ricci scalar replaced by the non-metricity that defines the gravity symmetric teleparallel formulation. In $f(Q, T)$ theory, the coupling between $Q$ and $T$ contributes to the non-conservation of the energy-momentum tensor, similar to the regular curvature trace of the energy-momentum tensor couplings. Many studies on couplings between matter and non-metricity are also conducted. T. Harko et al.\cite{T/2018} studied coupling matter in modified $Q$ gravity, novel couplings between nonmetricity and matter\cite{Tharko/2019}. Minimal coupling in the presence of non-metricity and torsion is investigated by A. Delhom\cite{Delhom/2020}. Thomas P Sotiriou\cite{Thomas/2009} also worked on $f(R)$ gravity, torsion and non-metricity.\\
 The $f(Q,T)$ gravity has been shown compatible with the accelerating expansion phase \cite{Yixin/2019}, besides it is also in agreement with important different phases our Universe passes through as the baryogenesis \cite{Sahoo/epjc_2020}. Moreover, such a theory is compatible with measurements of the Hubble parameter for different redshifts as one can see in  \cite{Simran/2020}. Beyond these successful tests, an important role any alternative theory of gravity should obey is the energy condition constraints \cite{Capozziello/2018}. These constraints are crucial to determine the proper regimes allowed for a new theory of gravity, to describe its attractive nature, and to assign the causal and the geodesic structure of space-time. Furthermore, the energy conditions also allow us to confront a new theory of gravity against $\Lambda$CDM model.

Therefore, in this work, we intend to study carefully all the energy conditions constraints on different forms of $f(Q,T)$ gravity. Our analyses were carried using the actual values of the Hubble, and the deceleration parameters. The energy conditions enable us to impose constraints over our free parameters, unveiling the viability of the $f(Q,T)$ models. We also verify the compatibility of our results with $\Lambda$CDM model. The discussions along this study are organized in the following nutshell: in section \ref{sec2} we introduce generalities about the $f(Q,T)$ gravity. In section \ref{sec3} we use the Raychaudhuri equations to find our energy conditions embedding the non-metricity and the trace of the energy-momentum tensor contributions. The constraints on $f(Q,T)$ models are discussed in details in section \ref{sec4}. A comparison between the $f(Q,T)$ models and the $\Lambda$CDM model is presented in section \ref{sec5}, where we also depicted the equation of state parameters for the models here studied. Section \ref{sec6} is dedicated to our final remarks and perspectives. 

\section{Overview of $f(Q, T)$ Gravity}\label{sec2}

The $f(Q,T)$ gravity is described through the following action \cite{Yixin/2019},

\begin{equation}  \label{1}
S=\int \left( \frac{1}{16\pi} f(Q,T) + L_{m}\right)d^4x \sqrt{-g} .
\end{equation}
where $f$ is an arbitrary function of the non-metricity $Q$, and of the trace of the energy-momentum tensor $T$, besides $L_{m}$ represents the Lagrangian of a given matter, and $g = det(g_{\alpha \beta})$. As it was discussed by Jimenez et al. \cite{Jimenez/2018}, the non-metricity function is such that

\begin{equation}  \label{2}
Q\equiv -g^{\alpha \beta}(L^{\mu}_{\,\, \nu \alpha}L^{\nu}_{\,\,\beta \mu}-L^{\mu}_{\,\,\nu \mu}L^{\nu}_{\,\,\alpha \beta}),
\end{equation}
where $L^{\mu}_{\,\, \nu\gamma}$ is the disformation tensor whose explicit form is 
\begin{equation}  \label{3}
L^{\mu}_{\, \, \nu\gamma}=-\frac{1}{2}g^{\mu\lambda}(\nabla_{\gamma}g_{\nu\lambda}+\nabla_{\nu}g_{\lambda\gamma}-\nabla_{\lambda}g_{\nu\gamma}).
\end{equation}

Another key ingredient to describe the symmetric teleparallel is the non-metricity tensor, which is defined as
\begin{equation}
\label{4}
Q_{\gamma\mu\nu}=\nabla_{\gamma}g_{\mu\nu}\,,
\end{equation}
and whose traces are
\begin{equation}
\label{5}
Q_{\mu}={{Q_{\mu}}^{\alpha}}_{\alpha}\,; \qquad \widetilde{Q}_{\mu}=Q^{\alpha}_{\,\alpha\,\mu}\,.
\end{equation}
 Here $Q_{\mu}$ is refereed as the Weyl vector, while $\widetilde{Q}_{\mu}$ is called the second non-metricity vector. 
We can also define a superpotential related with the non-metricity tensor as
\begin{equation}
\label{6}
4{P^{\mu}}_{\alpha\beta}=-{Q^{\mu}}_{\alpha\beta}+ 2{Q_{(\alpha}}^{\, \mu}_{\,\,\,\beta)}-Q^{\mu}g_{\alpha\beta}-\widetilde{Q}^{\mu}g_{\alpha\beta}-\delta^{\mu}_{(\alpha} Q_{\beta)}\,,
\end{equation}
yielding to the quadratic form for the non-metricity function \cite{Jimenez/2018}
\begin{equation}
\label{7}
Q=-Q_{\mu\alpha\beta}P^{\mu\alpha\beta}\,.
\end{equation}

Moreover, as it is known the energy-momentum tensor can be written as
\begin{equation}  \label{8}
T_{\alpha \beta}= -\frac{2}{\sqrt{-g}} \dfrac{\delta(\sqrt{-g}L_{m})}{\delta g^{\alpha \beta}}
\end{equation} 
and  its variation in respect to the metric tensor is such that
\begin{equation} \label{9}
\frac{\delta\,g^{\,\mu\nu}\,T_{\,\mu\nu}}{\delta\,g^{\,\alpha\,\beta}}= T_{\,\alpha\beta}+\Theta_{\,\alpha\,\beta}\,,
\end{equation}
where
\begin{equation}   \label{10}
\Theta_{\alpha \beta}= g^{\mu \nu} \frac{\delta T_{\mu \nu}}{\delta g^{\alpha \beta}}.
\end{equation}

Therefore, taking the variation of action \eqref{1} with respect to the metric, we find the field equations
\begin{multline}  \label{11}
8\pi T_{\alpha \beta}= -\frac{2}{\sqrt{-g}}\nabla_{\mu}(f_{Q}\sqrt{-g} P^{\mu}_{\,\, \alpha\beta}-\frac{1}{2}fg_{\alpha \beta}\\
+ f_{T}(T_{\alpha \beta}+\Theta_{\alpha \beta})-f_{Q}(P_{\alpha\mu \nu}Q_{\beta}^{\,\,\,\mu \nu}-2Q^{\mu\nu}_{\, \, \, \alpha}P_{\mu\nu \beta}).
\end{multline}
where $f_{Q}= \dfrac{df}{dQ}$.

Now, let us assume a Friedmann-Lemaitre-Robertson-Walker metric (FLRW) given by, 
\begin{equation}  
ds^{2}= -N^{2}(t)dt^{2}+ a^{2}(t)\delta_{ij}dx^{i}dx^{j},
\end{equation}
 where $a(t)$ is the scale factor of the Universe and $N(t)$ is the lapse function. Jim\'{e}nez et al. \cite{Jose} stated that if we choose the coincident gauge $\Gamma^{\alpha}_{\,\,\,\mu \nu}=0$, the diffeomorphism gauge's symmetry gets compromised in the generic theory. Hence, we cannot arbitrarily choose the time parameterization as used in General Relativity. The non-metricity scalar is defined to be $Q=6\frac{H^{2}}{N^{2}}$. As we used diffeomorphism to repair the coincident gauge, we cannot choose any unique lapse function. However, as $Q$ maintains a residual time reparametrization invariance stated in \cite{Jimenez/2018}, we are allowed to set $N=1$, as a matter of simplicity. Therefore, we yield to the metric,
\begin{equation}  \label{12}
ds^{2}= -dt^{2}+ a^{2}(t)\delta_{ij}dx^{i}dx^{j}\,.
\end{equation}
Furthermore, the matter content of the Universe is assumed as been a perfect fluid, whose energy-momentum tensor is $T_{\,\alpha\beta} = diag(-\rho, p, p, p)$. 

Therefore, substituting Eqs. \eqref{7}, \eqref{8}, and  \eqref{10} into  \eqref{9}, we yield to the modified Friedmann equations for such a theory, which are explicitly represented as
\begin{equation} \label{13}
3\,H^{\,2}=\frac{1}{2\,F}\,\left(-8\,\pi\,\rho+\frac{f}{2}-\frac{2\,\overline{G}}{1+\overline{G}}\,\left(\dot{F}\,H+F\,\dot{H}\right)\right)\,,
\end{equation}
\begin{equation}\label{14}
\dot{H}+3\,H^{\,2}=\frac{8\,\pi\,p}{2\,F}+\frac{f}{4\,F}-\frac{\dot{F}}{F}\,H\,.
\end{equation}
From the Friedmann equations, one can prove that the density and pressure also satisfy the generalized energy balance equation for $f(Q,T)$ gravity \cite{Yixin/2019}, whose form is
\begin{equation} \label{ge01}
\dot{\rho}+3\,H\,(\rho+p)= B_\mu\,v^{\mu}\,,
\end{equation}
where
\begin{eqnarray}
B_{\mu}v^{\mu}& =& \frac{\overline{G}}{16\,\pi\,\left(1+\overline{G}\right)\,\left(1+2\overline{G}\right)} \\ \nonumber
&&
\times\left[\dot{S}-\frac{\left(3\overline{G}+2\right)\dot{\overline{G}}}{\left(1+\overline{G}\right)\overline{G}}\,S+6\,H\,S\right]\,,
\end{eqnarray}
with $S = 2\,\left(\dot{F}\,H+\dot{H}\,F\right)$\,.

The previous equations can be rewritten in analogy to Einstein's general relativity in the following way
\begin{equation} \label{15}
3\,H^{\,2}=-\frac{8\,\pi\,}{2}\,\rho_{eff}\,,
\end{equation}
\begin{equation} \label{16}
\dot{H}+3\,H^{\,2}=\frac{8\,\pi}{2}\,p_{eff}\,,
\end{equation}
resulting in 
\begin{equation}\label{17}
\rho_{eff}=\frac{\rho}{F}-\frac{f}{16\,\pi\,F}+\frac{\overline{G}}{1+\overline{G}}\,\frac{\dot{F}\,H+F\,\dot{H}}{4\,\pi\,F}\,,
\end{equation}
\begin{equation}\label{18}
p_{eff}=\frac{p}{F}+\frac{f}{16\,\pi\,F}-\frac{\dot{F}}{4\,\pi\,F}\,H\,,
\end{equation}
as the effective density, and pressure. Here ($\cdot$) dot represents a derivative with respect to time, besides $F= f_{Q}$, and $8 \pi \overline{G}=f_{T}$ denote differentiation with respect to $Q$, and $T$, respectively. Moreover, we are able to observe that the contributions coming from the $f(Q,T)$ model are embedded into $\rho_{eff}$, and $p_{eff}$. We also highlight that this same identification procedure was applied to generalized Gauss-Bonnet gravity by Kaczmarek et. al. \cite{Kaczmarek/2020}, and also in $f(Q)$ gravity \cite{Mandal/2020}. Another interesting feature related with $\rho_{eff}$, and $p_{eff}$ is that they obey the conservation equation
\begin{equation} \label{eq_cons}
\dot{\rho}_{eff}+3\,H\,\left(\rho_{eff}+p_{eff}\right)=0\,.
\end{equation}
Such a behavior is going to be used in the next section, to address the physical interpretations related with the different energy conditions.

\section{Weyl-type $f(Q,T)$ and Raychaudhuri Equation} \label{sec3}

Energy conditions in modified gravity are the tools which empower the casual and geodesic structure of space-time. These conditions are formulated with the help of Raychaudhuri equations which describe the action of congruence and attractiveness of the gravity for timelike, spacelike, or lightlike curves. 

In order to study the implications of the non-metricity, and of the trace of the energy-momentum tensor for the Raychaudhuri equation, we are going to derive such an equation in the Weyl framework. The Weyl geometry is based on a connection where the orientation and the magnitude of a vector can change, under a parallel transportation. Therefore, such a geometry is a natural framework to describe the effects of torsion and of the non-metricity on the particle dynamics. 

Recently, Yixin Xu et al. \cite{Yixin/2020} extended the formulation of $f(Q,T)$ in the framework of Weyl geometry. There, the authors were able to derive the general field equations for gravity in the Weyl geometry coupled to the matter energy-momentum tensor, as well as, to analyze the cosmological implications of different families of $f(Q,T)$ gravity. In another recent work Jin-Zhao Yang et al. \cite{Yang/2021}, complemented the formulation raised in \cite{Yixin/2020} and also in the beautiful work of Iosifidis et al. \cite{Damianos/2018}, studying the geodesic deviation equation, the Raychaudhuri equation, and the tidal forces for the $f(Q,T)$ gravity in Weyl framework.  In order to implement our set of energy conditions for the $f(Q,T)$ models introduced in \cite{Yixin/2019}, based on the constraints over the Raychaudhuri equation, let us briefly reproduce some definitions and results from \cite{Yixin/2020, Yang/2021, Damianos/2018}.

The Weyl geometry is a generalization of the Riemannian geometry, where the connection is redefined as
\begin{equation}
\widetilde{\Gamma}^{\lambda}_{\,\mu\,\nu}\equiv \Gamma^{\lambda}_{\,\mu\,\nu}+g_{\mu\nu} w^{\lambda}-\delta^{\lambda}_{\mu}w_{\nu}-\delta^{\lambda}_{\nu}w_{\mu}\,,
\end{equation}
where $w_{\mu}$ is the so-called Weyl vector, and $\Gamma^{\lambda}_{\,\mu\,\nu}$ is determined through the metric $g_{\mu\,\nu}$. Taking the Weyl connection, one is able to show that the covariant derivative applied to the metric tensor is

\begin{equation}
\widetilde{\nabla}_{\lambda}\,g_{\,\mu\nu}=2\,w_{\lambda}g_{\,\mu\nu}\,;\qquad \widetilde{\nabla}_{\lambda}\,g^{\,\mu\nu}=-2\,w_{\lambda}\,g^{\,mu\nu}\,.
\end{equation}
Therefore, in this framework the non-metricity tensors are written as
\begin{eqnarray}
&&
Q_{\,\lambda\mu\nu} = -\widetilde{\nabla}_{\lambda}\,g_{\,\mu\nu}=-2\,w_{\lambda}g_{\,\mu\nu}\,; \\ \nonumber
&&
Q^{\,\lambda\mu\nu} = \widetilde{\nabla}^{\lambda}\,g^{\,\mu\nu}=-2\,w^{\lambda}g^{\,\mu\nu}\,.
\end{eqnarray}
Such relations also yield to 
\begin{equation}
Q = -6 w^2\,,
\end{equation}
as the non-metricity scalar. 

The presence of the non-metricity affects the length of vectors in Weyl framework, when they are submitted to parallel transport. In such a case, the four velocity is given by
\begin{equation}
u_{\mu}\,u^{\mu}= g_{\mu\nu} u^{\,\mu} u^{\,\nu}=-l^2\,, \qquad l =l (x^{\alpha})\,,
\end{equation}
where $u=dx^{\mu}/d\lambda$ with $\lambda$ as the affine parameter, and $l(x^{\alpha})$ is an arbitrary function of space and time coordinates. Moreover, in the presence of the non-metricity the associated projection tensor has the form
\begin{equation}
h_{\mu\nu}=g_{\mu\nu}+\frac{1}{l^2}\,u_{\mu}\,u_{\nu}\,,
\end{equation}
which results in the following properties:
\begin{eqnarray}
&&
h_{\mu\nu} = h_{\nu\mu}\,; \\ \nonumber
&&
h_{\mu\nu}u^{\mu}=0\,; \\ \nonumber
&&
h_{\mu\nu}h^{\mu\nu}=3 \\ \nonumber
&&
h_{\mu\lambda}h^{\lambda\nu}=h_{\mu}^{\nu}= \delta_{\mu}^{\nu}+\frac{1}{l^2}u_{\mu}u^{\nu}\,.
\end{eqnarray}
Besides, the non-metricity also implies in two different acceleration vectors, since raising and lowering indexes do not commute. Such acceleration vectors are defined as \cite{Damianos/2018}
\begin{equation} \label{acc_01}
A^{\mu}=u^{\prime \, \mu} = u^{\lambda}\widetilde{\nabla}_{\lambda}u^{\mu}\,,
\end{equation}
as the contravariant or 4-acceleration vector, and
\begin{equation} \label{acc_02}
a_{\mu}=u^{\prime}_{\,\mu}=u^{\lambda}\widetilde{\nabla}_{\lambda}u_{\mu}\,,
\end{equation}
as the covariant or hyper 4-acceleration vector. In these last equations, prime means a derivative in respect to the affine parameter $\lambda$. These definitions yield to the following constraint between the acceleration vectors,
\begin{equation}
A^{\mu}=a^{\mu}+Q^{\,\nu\lambda\mu}u_{\nu}u_{\lambda}\,.
\end{equation}

From Eqs. \eqref{acc_01} and \eqref{acc_02}, we are able to derive that
\begin{equation}
A^{\mu}u_{\mu}=-\frac{(l^2)^{\prime}}{2}+\frac{1}{2}\,Q_{\mu\nu\lambda}u^{\mu}u^{\nu}u^{\lambda}\,,
\end{equation}
and
\begin{equation}
a^{\mu}u_{\mu} = -\frac{(l^2)^{\prime}}{2}-\frac{1}{2}\,Q_{\mu\nu\lambda}u^{\mu}u^{\nu}u^{\lambda}\,.
\end{equation}
Therefore, we directly observe that
\begin{equation}
\left(A^{\mu}+a^{\mu}\right)u_{\mu} = - (l^2)^{\prime}\,;\qquad \left(A^{\mu}-a^{\mu}\right)u_{\mu} =  Q_{\mu\nu\lambda}u^{\mu}u^{\nu}u^{\lambda}\,.
\end{equation}

Another key ingredients to build the Raychaudhuri equation in this framework are the covariant derivatives of $u_{\mu}$ and $a_{\mu}$ vectors, whose temporal and spacial components are
\begin{equation}
\widetilde{\nabla}_{\nu}u_{\mu}={\cal D}_{\nu}u_{\mu}-\frac{1}{l^2}\left(u_{\mu}\xi_{\nu}+a_{\mu}u_{\nu}\right)-\frac{1}{l^4}\left(u^{\lambda}a_{\lambda}\right)u_{\mu}u_{\nu}\,,
\end{equation}
\begin{equation}
\widetilde{\nabla}^{\mu}a_{\mu}= {\cal D}^{\mu}a_{\mu}+\frac{1}{l^2}\,A^{\mu}a_{\mu}-\frac{1}{l^2}a_{\mu} u^{\mu}\,,
\end{equation}
where ${\cal D}_{\nu}u_{\mu}=h_{\nu}^{\varphi}h_{\mu}^{\lambda}\widetilde{\nabla}_{\varphi}u_{\lambda}$ is the projected covariant derivative, besides $\xi_{\mu}=u^{\nu}\widetilde{\nabla}_{\mu}u_{\nu}$, and $\xi_{\mu}u^{\mu}=a_{\mu}u^{\mu}$. The projected covariant derivative of vector $u_{\mu}$ can be decomposed as
\begin{equation}
D_{\nu}u_{\mu} = \frac{1}{3}\,\left(\theta+\frac{1}{l^2}a_{\lambda}u^{\lambda}\right)\,h_{\mu\nu}+\sigma_{\mu\nu}+\omega_{\mu\nu}\,,
\end{equation}
for $\theta$, $\sigma_{\mu \nu}$, and $\omega_{\mu \nu}$ as the expansion, shear, and rotation, associated to the vector field $u^{\mu}$, respectively.

The previous definitions allow us to write the following geodesic equation
\begin{equation}
A^{\mu}=\frac{d^2x^{\mu}}{d\lambda^2}+\widetilde{\Gamma}^{\mu}_{\nu\lambda}u^{\nu}u^{\lambda}=f^{\mu}\,,
\end{equation}
where $f^{\mu}$ is interpreted as an extra force induced by the non-minimal coupling between $Q$ and $T$ at the $f(Q,T)$ gravity. 

The Raychaudhuri equation in this framework can be derived from the curvature tensor
\begin{equation}
\left(\widetilde{\nabla}_{\mu}\widetilde{\nabla}_{\nu}-\widetilde{\nabla}_{\nu}\widetilde{\nabla}_{\mu}\right)u_{\lambda}=-\widetilde{R}_{\beta\lambda\nu\mu}u^{\beta}\,,
\end{equation}
where we are not considering torsion effects. Contracting the last equation with $g^{\lambda\nu}u_{\mu}$ yields to
\begin{equation}
g^{\lambda\nu}u_{\mu}\left(\widetilde{\nabla}_{\mu}\widetilde{\nabla}_{\nu}-\widetilde{\nabla}_{\nu}\widetilde{\nabla}_{\mu}\right)u_{\lambda}=-\widetilde{R}_{\beta\lambda\nu\mu}u^{\beta}u^{\mu}g^{\lambda\nu}\,.
\end{equation}
With these relations in hand, one can determine that the Raychaudhuri equation is explicitly written as
\begin{eqnarray} \label{ray01}
&&
\theta^{\prime} = -\frac{\theta^2}{3}-R_{\mu\nu} u^{\mu} u^{\nu}-\sigma_{\mu\nu}\sigma^{\mu\nu}+\omega_{\mu\nu}\omega^{\mu\nu} \\ \nonumber
&&
+\widetilde{\nabla}^{\mu}a_{\mu}-\frac{2}{3\,l^2}\theta\,a_{\mu}u^{\mu}+\frac{2}{3\,l^4}\left(a_{\mu}u^{\mu}\right)^2+\frac{2}{l^2}a_{\mu}\xi^{\mu}-\widetilde{Q}^{\prime}_{\mu}u^{\mu} \\ \nonumber
&&
+\frac{1}{3}\left(\theta+\frac{1}{l^2}a_{\nu}u^{\nu}\right)\left(Q_{\mu}-\widetilde{Q}_{\mu}\right)\,u^{\mu}-Q_{\mu\nu\lambda}\left(\sigma^{\mu\nu}+\omega^{\mu\nu}\right)u^{\lambda} \\ \nonumber
&&
-\frac{1}{l^2}\,Q_{\mu\nu\lambda}u^{\mu}u^{\nu}\left(a^{\lambda}+\xi^{\lambda}\right)+Q_{\mu\nu\lambda}u^{\mu}\sigma^{\nu\lambda}\\ \nonumber
&&
+\frac{1}{l^2}Q_{\mu\nu\lambda}\left(u^{\mu}\xi^{\nu}+a^{\mu}u^{\nu}\right)u^{\lambda}+u^{\mu}u^{\nu}\widetilde{\nabla}^{\lambda}Q_{\mu\nu\lambda}+Q_{\mu}^{\lambda\beta}Q_{\beta\lambda\nu}u^{\mu}u^{\nu}\,.
\end{eqnarray}
The detailed calculation of the previous equation can be found in Refs. \cite{Yang/2021, Damianos/2018}. Now, taking the Weyl and the second non-metricity vectors from Eq. \eqref{5}, and constraining the Raychaudhuri equation for autoparallel curves, where particle have zero path acceleration $\left(A^{\mu}=f^{\mu}=0\right)$ \cite{Damianos/2018}, Eq. \eqref{ray01} is reduced to
\begin{equation}\label{ray02}
\left(\theta-2\,\frac{l^{\prime}}{l}\right)^{\prime} = -\frac{1}{3}\,\left(\theta-2\,\frac{l^{\prime}}{l}\right)^{2}
-R_{\mu\nu}u^{\mu}u^{\nu}-2\,\left(\sigma^2-\omega^2\right)\,.
\end{equation}

Then, following the procedure introduced by Iosifidis et al. \cite{Damianos/2018}, we can assume that for an irrotational and shear-free scenario, \eqref{ray02} yields to the constraint
\begin{equation}
\left(\theta-2\,\frac{l^{\prime}}{l}\right)^{\prime}+\frac{1}{3}\,\left(\theta-2\,\frac{l^{\prime}}{l}\right)^{2}\leq 0\,,
\end{equation}
if 
\begin{equation} \label{21}
R_{\mu \nu} u^{\mu} u^{\nu} \geqslant 0\,,
\end{equation}
resulting in an attractive behavior for gravity. Therefore, this approach unveils a generalization of the constraints for the energy conditions. Note that by changing $\theta-2\,\frac{l^{\prime}}{l} \rightarrow \theta$ into \eqref{ray02}, we can recover the standard Raychaudhuri equation for GR. Moreover, the same procedure here discussed can be applied to the case of a null-vector $k^{\mu}$, leading to the following reduced form of the Raychaudhuri equation 
\begin{equation}\label{ray03}
\left(\theta-2\,\frac{l^{\prime}}{l}\right)^{\prime} = -\frac{1}{3}\,\left(\theta-2\,\frac{l^{\prime}}{l}\right)^{2}
-R_{\mu\nu}k^{\mu}k^{\nu}-2\,\left(\sigma^2-\omega^2\right)\,.
\end{equation}
Consequently, we are able to establish the constraint
\begin{equation}
\left(\theta-2\,\frac{l^{\prime}}{l}\right)^{\prime}+\frac{1}{3}\,\left(\theta-2\,\frac{l^{\prime}}{l}\right)^{2}\leq 0\,,
\end{equation}
for 
\begin{equation} \label{22}
R_{\mu \nu} k^{\mu} k^{\nu} \geqslant 0\,,
\end{equation}
generalizing the approach of Santos et al. \cite{Santos}.

Following the methodology presented in \cite{Mandal/2020}, the $f(Q,T)$ models are going to be restricted to the set of energy conditions bellow
\begin{itemize}
\item null energy condition (NEC) $\Leftrightarrow \rho_{eff} + p_{eff} \geq 0$;
\item weak energy condition (WEC) $\Leftrightarrow
\rho_{eff} + p_{eff} \geq 0$ and $\rho_{eff} \geq 0$;
\item dominant energy condition (DEC) $\Leftrightarrow \rho_{eff} \geq |p_{eff}|$ and $\rho_{eff} \geq 0$;
\item strong energy condition (SEC) $\Leftrightarrow \rho_{eff}+ 3\, p_{eff} \geq 0$.
\end{itemize}

Through Eq. \eqref{eq_cons}, we are able to observe an expanding behavior for the Universe where energy density always decreases, if the null energy condition is obeyed \cite{Rubakov/2014}. Such a behavior is also corroborated by weak energy condition which also establishes that the local energy density should be always positive, which means that local observers measure positive mass densities \cite{Tipler/1978}. Beyond the expanding behavior constrained by WEC, the dominant energy condition also imposes that the pressure cannot exceed energy density, in order to state that matter flows along timelike or null world lines \cite{Capozziello/2018}. Moreover, when we deal with a perfect fluid, in this case mapped by $\rho_{eff}$, and $p_{eff}$, the strong energy condition states that we must have, besides $\rho_{eff} + p_{eff} \geq 0$, the sum $\rho_{eff} + 3\,p_{eff} \geq 0$, which means that SEC includes NEC and excludes large negative pressures regimes \cite{Crislane/2017}. Therefore, in order to describe a Universe that is dominated by negative pressure, the SEC condition needs to be violated \cite{Capozziello/2018}. 

Then, by substituting Eqs. \eqref{11}, and \eqref{12} in the previous relations, we established the following set of energy conditions
\begin{itemize}
\item null energy condition (NEC) $\Leftrightarrow \rho + p \geq 0$;
\item weak eneregy condition (WEC) $\Leftrightarrow
\rho + p \geq 0$;
\item dominant energy condition (DEC) $\Leftrightarrow \rho \geq |p|$.
\end{itemize}
Moreover, WEC, DEC and SEC energy conditions demand the extra constraints
\begin{equation} \label{23}
\mbox{DEC} \Leftrightarrow F\leq 0\,;
\end{equation}
\begin{equation} \label{24}
\mbox{WEC and DEC}\Leftrightarrow \rho-\frac{f}{16\,\pi}+\frac{\overline{G}}{1+\overline{G}}\,\frac{\dot{F}\,H+F\,\dot{H}}{4\,\pi}\geq 0\,,
\end{equation} 
and
\begin{equation} \label{25}
\mbox{SEC}\Leftrightarrow \rho+3\,p+\frac{f}{8\,\pi}-\frac{3-\overline{G}}{1+\overline{G}}\,\frac{\dot{F}\,H}{4\,\pi}+\frac{\overline{G}}{1+\overline{G}}\,\frac{F\,\dot{H}}{4\,\pi}\geq 0\,.
\end{equation}
Through these constraints, we can realize how different $f(Q,T)$ models modify the standard energy conditions derived from the Raychaudhuri equations.

\section{Constraining $f(Q,T)$ Gravity Models}\label{sec4}

In the framework of FRW metric, one can also use the constraints of energy conditions to restrict certain models in $f(Q, T)$ gravity. A cosmological quantity which is essential to properly describe the energy conditions in a phenomenological perspective is the deceleration parameter, whose definition is \cite{Sharif}

\begin{equation} \label{26}
q= -\frac{1}{H^{2}} \frac{\ddot{a}}{a}\,.
\end{equation}
Alternatively, the time derivative of the Hubble parameter can be rewritten as
\begin{equation} \label{29}
\dot{H}= -H^{2}(1+q).
\end{equation}
Beyond these ingredients, in order to constraint the energy conditions with phenomenological observations, we are going to consider that  $H=H_{0}=67.9 \,km\,s^{\,-1}\,Mpc^{\,-1}$, and $q=q_{0}=-0.503$ as the present values for the Hubble, and the deceleration parameters, respectively \cite{Planck/2018,Capozziello/2019}. 

\subsection{$f(Q, T)= m Q+ b T$}\label{A}

As a first model, we are going to work with $f (Q, T )=m Q + b T$, where $m$, and $b$ are free parameters. This model was introduced by Xu et al. \cite{Yixin/2019}, and it naturally describes an exponential expanding Universe, with $\rho \propto e^{-H_0\,t}$ \cite{Yixin/2019}. The present model yields to  $F=m$ and  $8\pi \overline{G}=b$, besides, by taking the modified Friedmann equations (\ref{13}), and (\ref{14}) together with Eqs. (\ref{26}), (\ref{29}) into the energy conditions, we find the constraints

\begin{widetext}
\begin{equation} \label{32}
NEC \Leftrightarrow -\frac{H_{0}^2 m (b (q_{0}+4)+24 \pi )}{2 \left(b^2+12 \pi  b+32 \pi ^2\right)}-\frac{H_{0}^2 m (3 b q_{0}+8 \pi  (2 q_{0}-1))}{2 \left(b^2+12 \pi  b+32 \pi ^2\right)} \geq 0\,,
\end{equation}

\begin{equation} \label{33}
WEC \Leftrightarrow -\frac{H_{0}^2 m (b (q_{0}+4)+24 \pi )}{2 \left(b^2+12 \pi  b+32 \pi ^2\right)}-\frac{H_{0}^2 m (3 b q_{0}+8 \pi  (2 q_{0}-1))}{2 \left(b^2+12 \pi  b+32 \pi ^2\right)}\geq 0\,,
\qquad and \qquad
-\frac{3 m H_0^2}{4 \pi } \geq 0\,,
\end{equation}

\begin{equation} \label{34}
DEC \Leftrightarrow \frac{H_{0}^2 m (3 b q_{0}+8 \pi  (2 q_{0}-1))}{2 \left(b^2+12 \pi  b+32 \pi ^2\right)}-\frac{H_{0}^2 m (b (q_{0}+4)+24 \pi )}{2 \left(b^2+12 \pi  b+32 \pi ^2\right)} \geq 0\,,
\qquad and \qquad
-\frac{3 m H_0^2}{4 \pi } \geq 0\,,
\end{equation}

\begin{equation}  \label{35}
SEC \Leftrightarrow (8 \pi )^{-1}\left( 6 H_{0}^2 m-\frac{2 b H_{0}^2 m (q_{0}+1)}{b+8 \pi }+\frac{2 H_{0}^2 m (-2 b q_{0}+b-24 \pi  q_{0})}{b+8 \pi }\right)  \geq 0.
\end{equation}

\begin{figure*}[h!]
\centering
\includegraphics[width=7.5 cm]{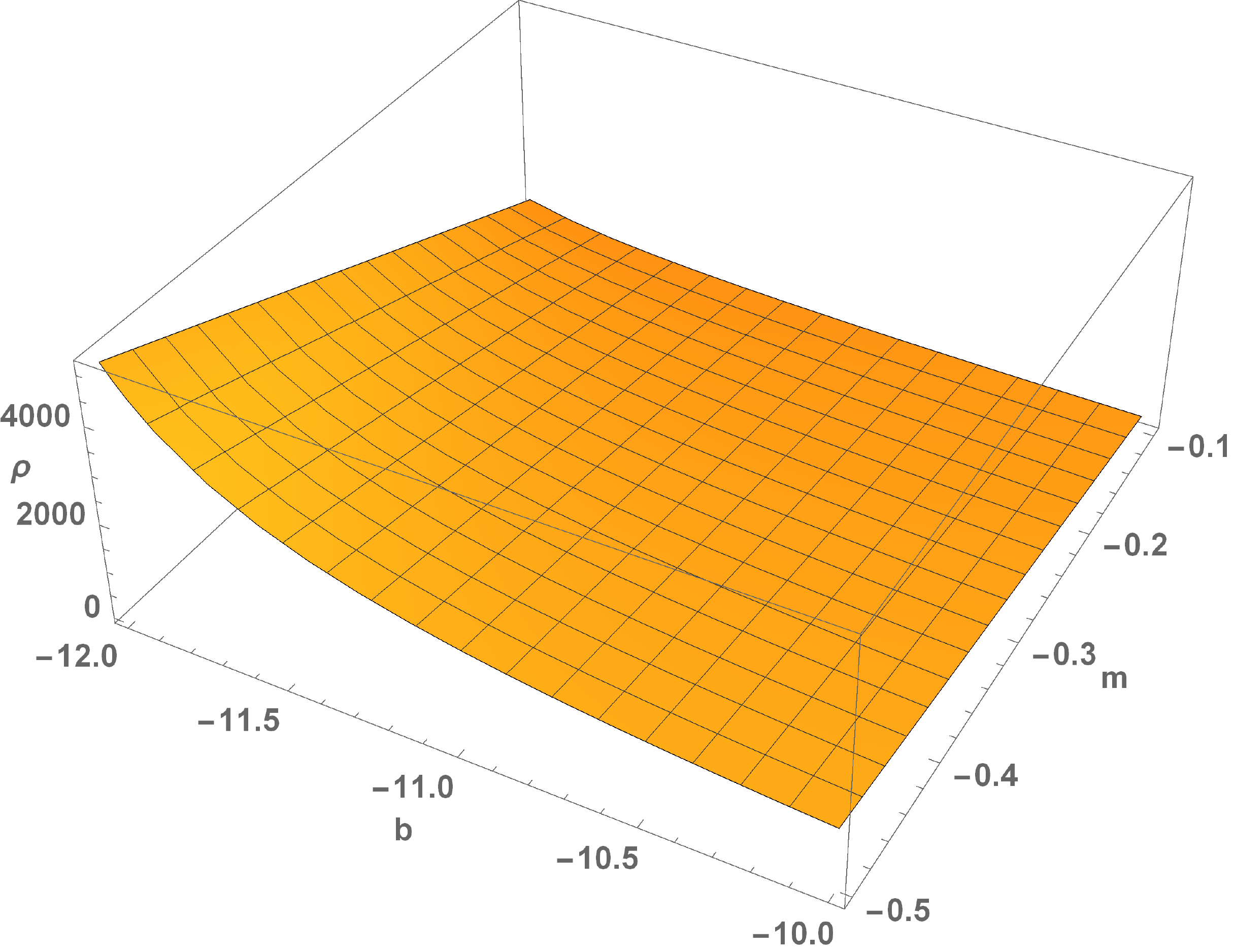} \hspace{0.5 cm}
\includegraphics[width=7.5 cm]{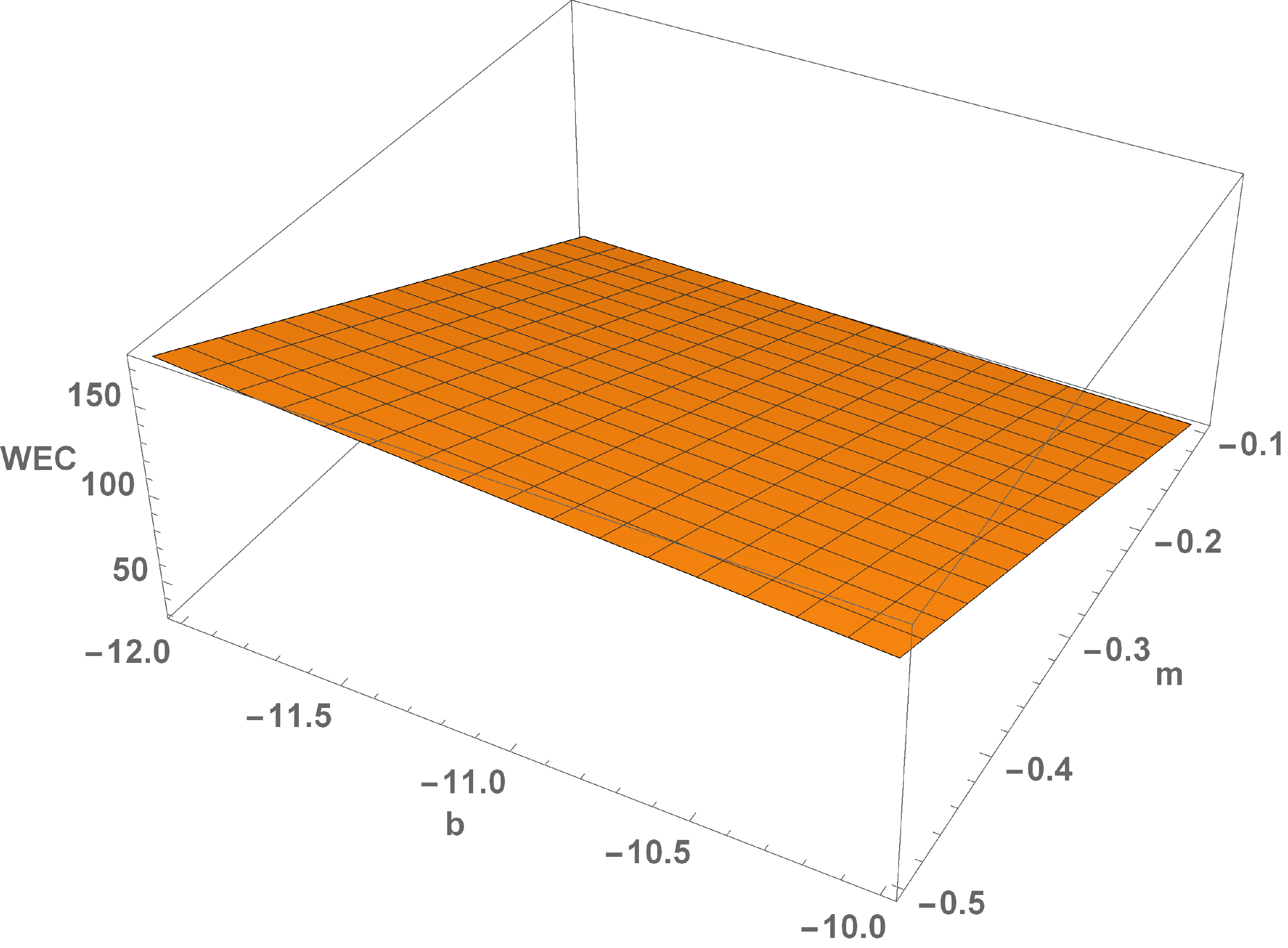} \hspace{0.5 cm}
\includegraphics[width=7.5 cm]{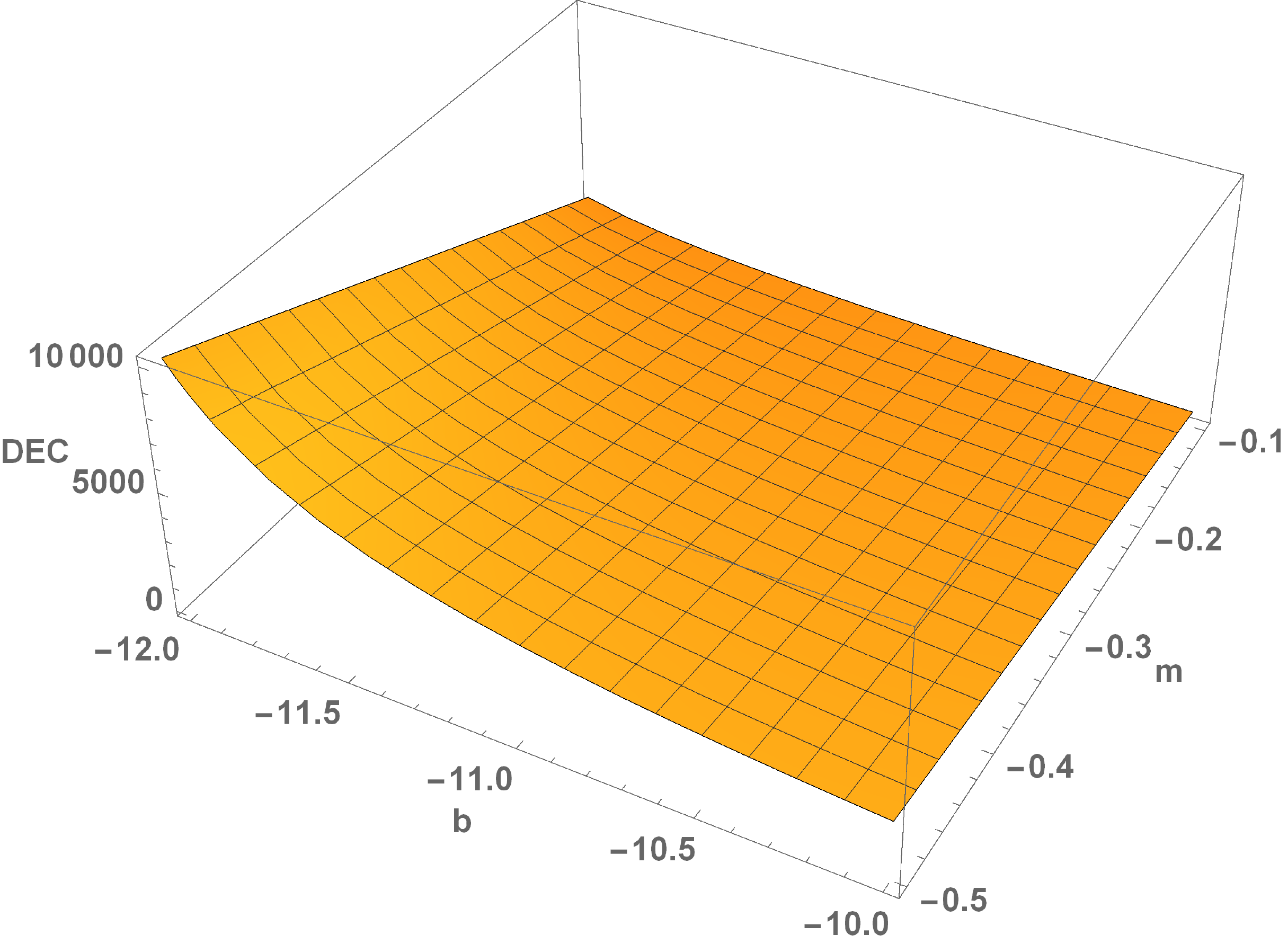} \hspace{0.5 cm}
\includegraphics[width=7.5 cm]{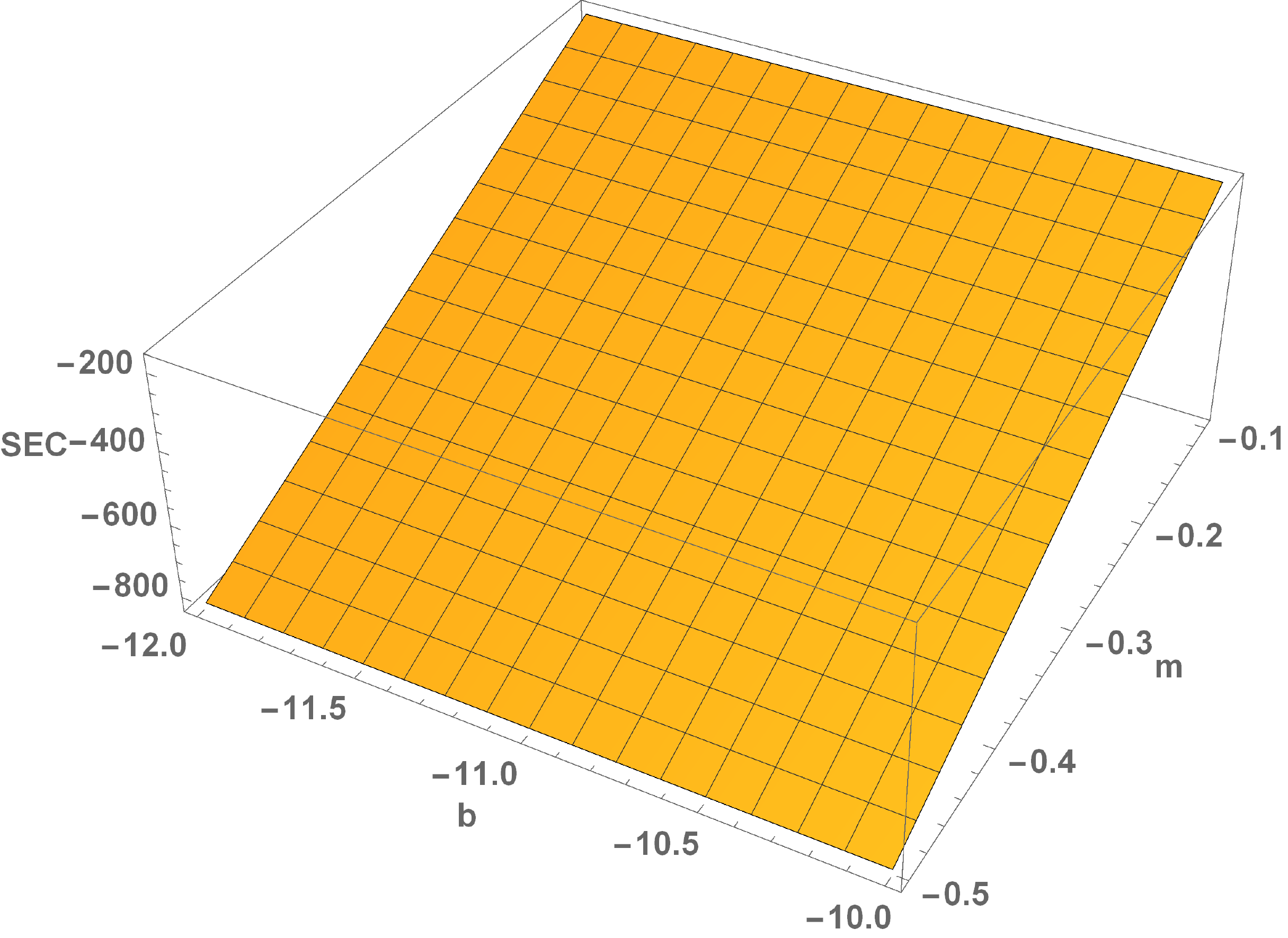}
\caption{Density parameter, and Energy conditions for $f(Q,T)= m Q+ b T$. The graphics were depicted with the present values of $H_0$ and $q_0$ parameters.}
\label{Fig-1}
\end{figure*}

\end{widetext}

We can observe the behavior of $\rho$, WEC, DEC, and SEC energy conditions in the graphics depicted in Fig. \ref{Fig-1}. There we realize how the density $\rho$ decreases for specific values of $m$, and $b$, corroborating with the exponential expansion behavior derived by Xu et al. \cite{Yixin/2019}. Moreover, the energy conditions allow us to constrain the free parameters $m$, and $b$. Through DEC we found that $m$ should be negative. Also \eqref{32}, \eqref{33}, and \eqref{35} assure the range of model parameters as $b > - 4 \,\pi$ and $m \leq 0$, satisfying NEC, WEC, and DEC. Furthermore, the constrained parameters result in the violation of SEC, which is compatible with the accelerated phase our Universe passes through \cite{Visser/2000}. Another remarkable feature coming from the energy conditions, is that the constrained parameters $m$, and $b$ corroborate, and fine tune the observational bounds for $f(Q,T)$ gravity investigated in \cite{Simran/2020}. 

\subsection{$f(Q, T)= Q^{n+1}+ b T$}\label{B}

As a second model, let us deal with  $f(Q,T)= Q^{n+1}+ b T$ where $n$ and $b$ are free parameters. Such a model was proposed by Xu et al. \cite{Yixin/2019}, and considers non-linear contributions due to the torsion in the gravity sector, moreover, it was constrained through measurements of the Hubble parameter for different redshifts \cite{Simran/2020}. This specific model also can describe an accelerating Universe with $\rho \propto t^{\,2\,(n-1)}$ as one can see in \cite{Yixin/2019}. The present model yields to $f =f_{Q}= (n+1)Q^{n}$ and $8\pi \overline{G} = f_{T} =b$, and by working with Eqs.  (\ref{13}), (\ref{14}), (\ref{26}), and (\ref{29}) we are able to derive the energy conditions below
\begin{widetext}
\begin{equation} \label{36}
NEC \Leftrightarrow -\frac{2^{n-1} 3^n (2 n+1) \left(H_{0}^2\right)^{n+1} (b (n q_{0}+n+q_{0}+4)+24 \pi )}{b^2+12 \pi  b+32 \pi ^2}-\frac{\splitdfrac{2^{n-1} 3^n (2 n+1) \left(H_{0}^2\right)^{n+1} (3 b (n q_{0}+n+q_{0})}{+8 \pi  (2 n (q_{0}+1)+2 q_{0}-1))}}{b^2+12 \pi  b+32 \pi ^2}\geq 0\,,
\end{equation}

\begin{multline} \label{37}
WEC \Leftrightarrow -\frac{2^{n-1} 3^n (2 n+1) \left(H_{0}^2\right)^{n+1} (b (n q_{0}+n+q_{0}+4)+24 \pi )}{b^2+12 \pi  b+32 \pi ^2}-\frac{\splitdfrac{2^{n-1} 3^n (2 n+1) \left(H_{0}^2\right)^{n+1} (3 b (n q_{0}+n+q_{0})}{+8 \pi  (2 n (q_{0}+1)+2 q_{0}-1))}}{b^2+12 \pi  b+32 \pi ^2} \geq 0\,,\\ 
\qquad and \qquad
-\frac{2^{n-2} 3^{n+1} (n+1) \left(H_0^2\right)^{n+1}}{\pi } \geq 0\,,
\end{multline}

\begin{multline} \label{38}
DEC \Leftrightarrow \frac{\splitdfrac{2^{n-1} 3^n (2 n+1) \left(H_{0}^2\right)^{n+1} (3 b (n q_{0}+n+q_{0})}{+8 \pi  (2 n (q_{0}+1)+2 q_{0}-1))}}{b^2+12 \pi  b+32 \pi ^2}-\frac{\splitdfrac{2^{n-1} 3^n (2 n+1) \left(H_{0}^2\right)^{n+1}}{ (b (n q_{0}+n+q_{0}+4)+24 \pi )}}{b^2+12 \pi  b+32 \pi ^2} \geq 0\,, \\
\qquad and \qquad
-\frac{2^{n-2} 3^{n+1} (n+1) \left(H_0^2\right)^{n+1}}{\pi } \geq 0\,,
\end{multline}

\begin{equation} \label{39}
SEC \Leftrightarrow -\frac{2^{n-2} 3^{n+1} (n+1) \left(H_0^2\right)^{n+1} (b (2 n (q_0+1)+q_0-1)+8 \pi  (q_0-1))}{\pi  (b+8 \pi )} \geq 0\,.
\end{equation}

\begin{figure*}[h!]
\centering
\includegraphics[width=7.5 cm]{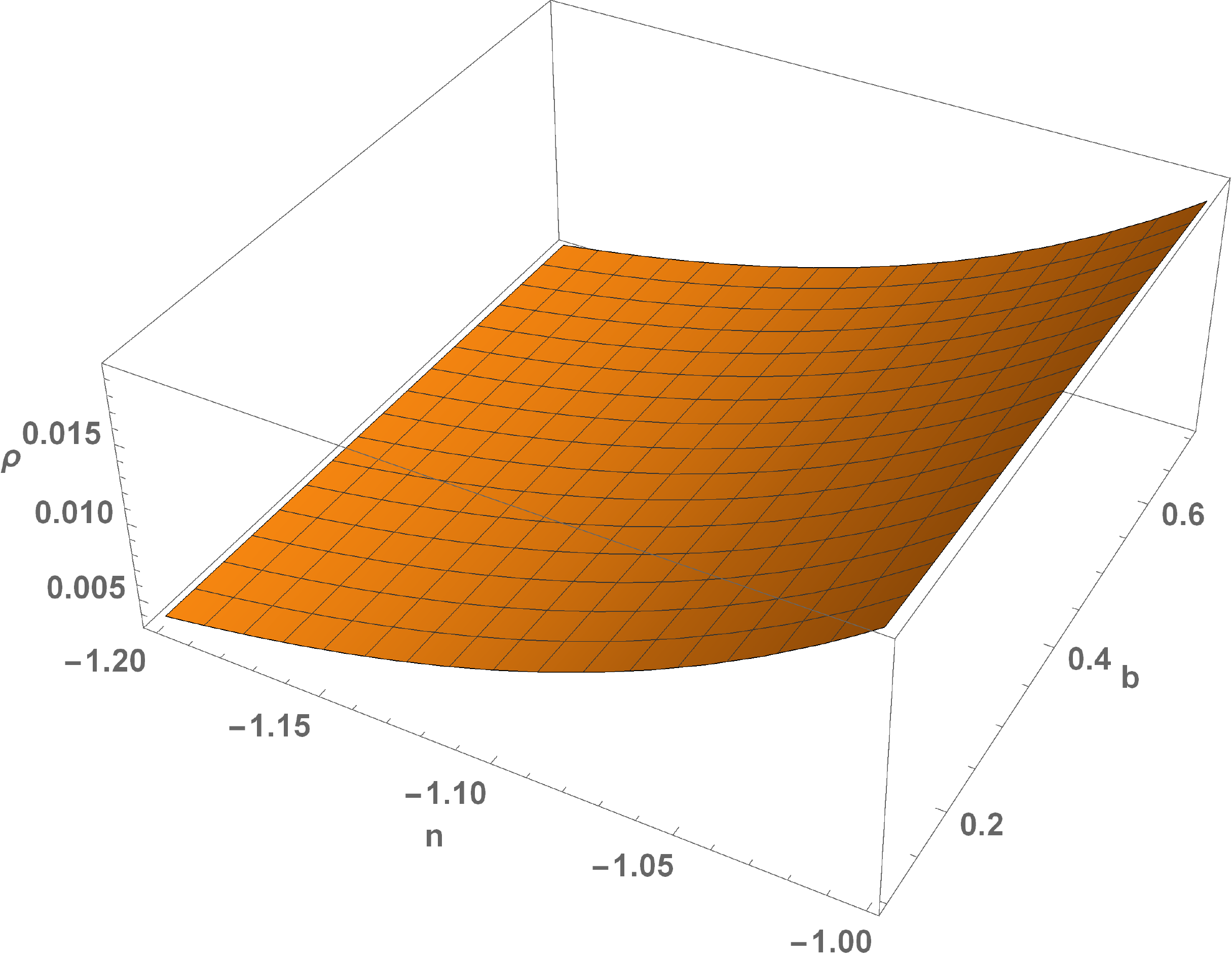} \hspace{0.5 cm}
\includegraphics[width=7.5 cm]{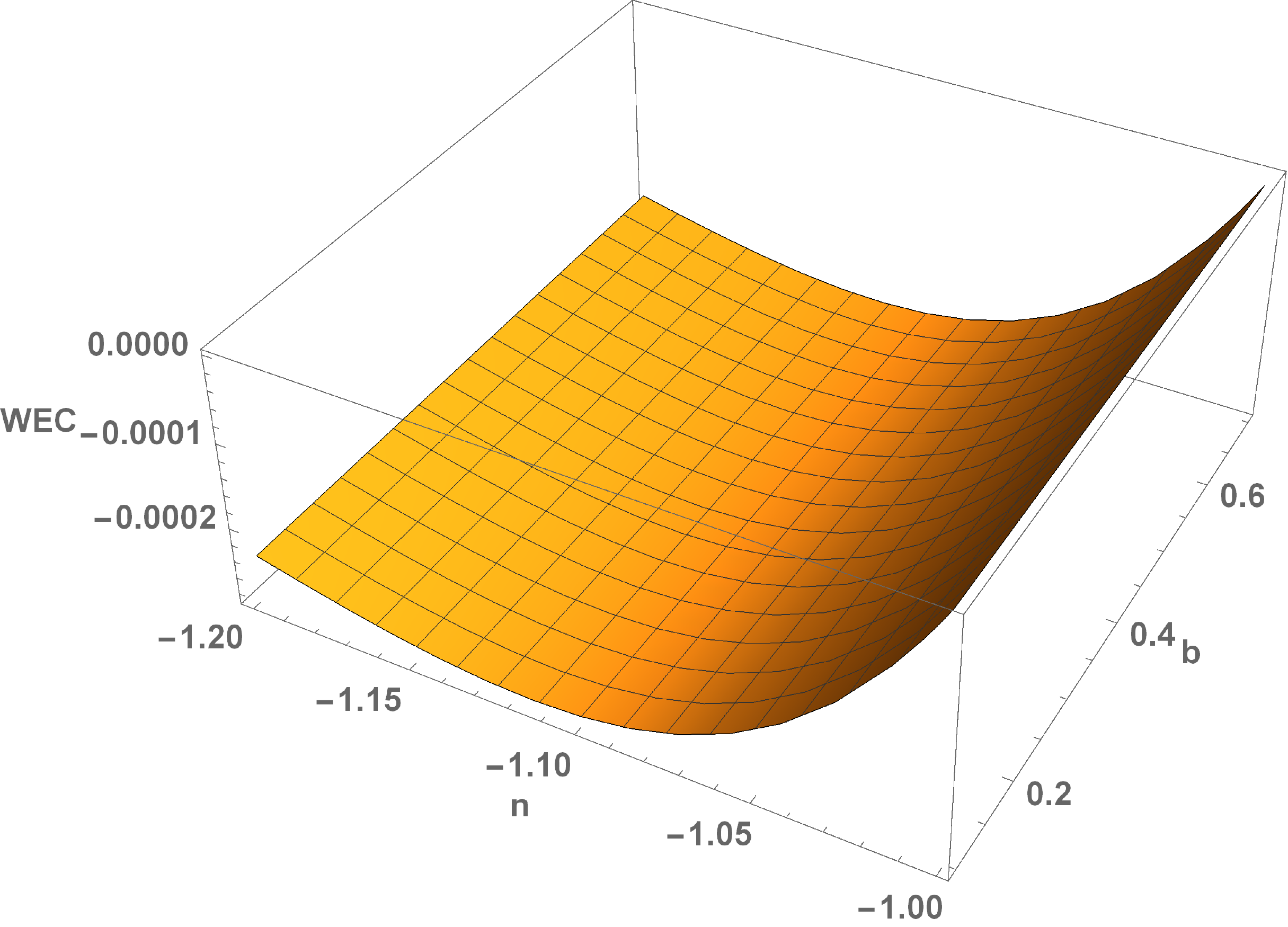} \hspace{0.5 cm}
\includegraphics[width=7.5 cm]{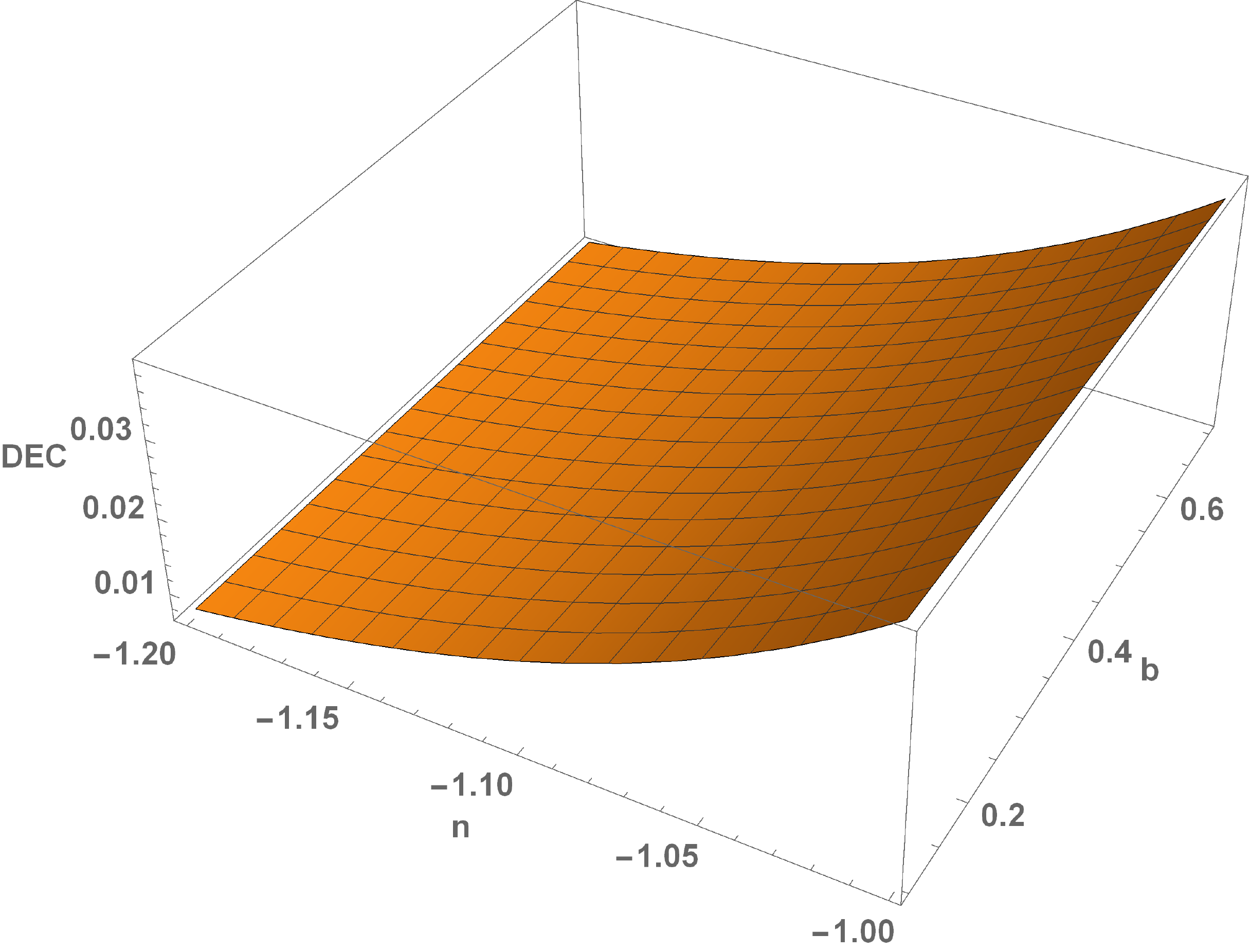} \hspace{0.5 cm}
\includegraphics[width=7.5 cm]{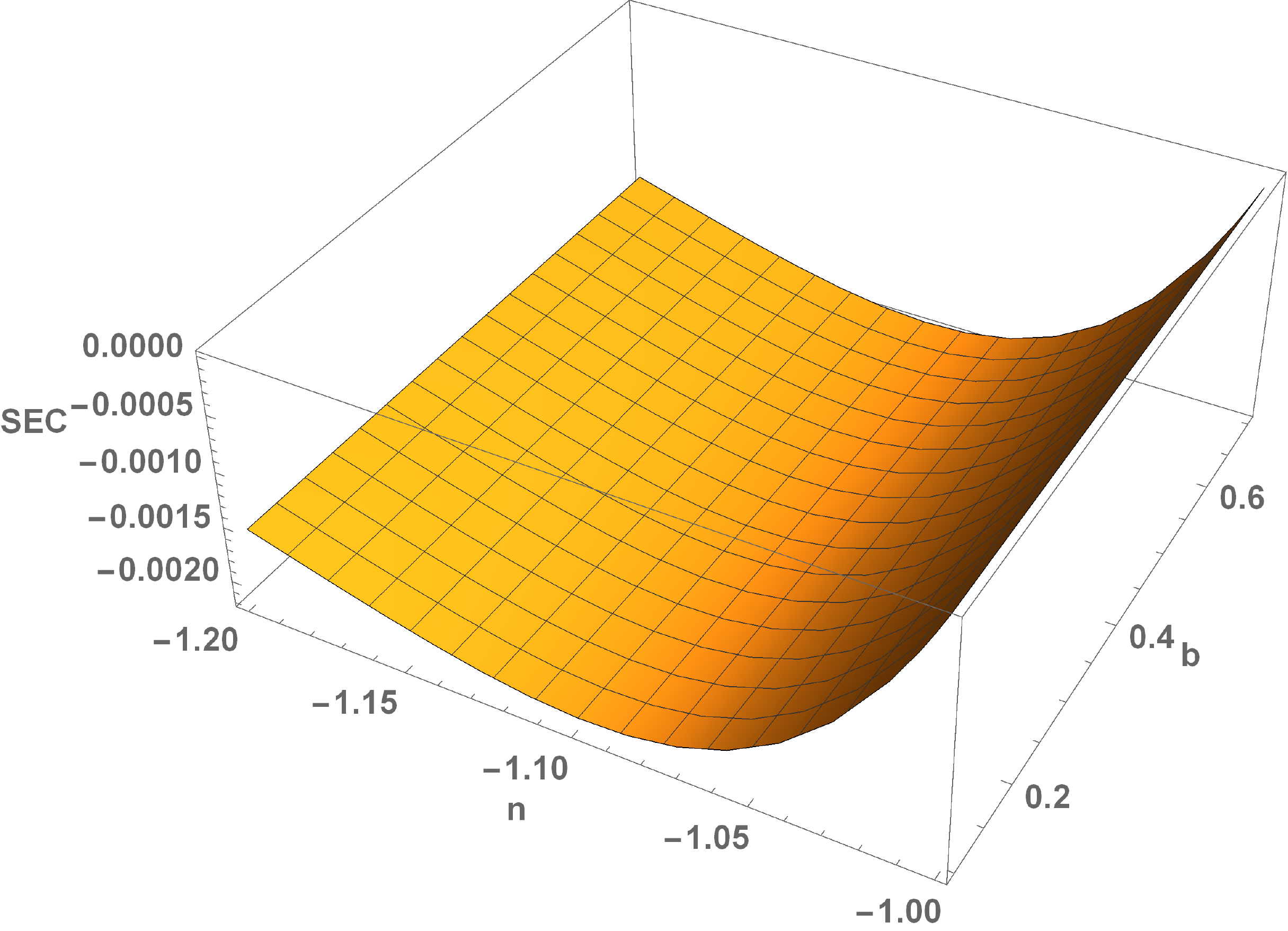}
\caption{Density parameter, and Energy conditions for $f(Q,T)= Q^{n+1}+ b T$. The graphics were depicted with the present values of $H_0$ and $q_0$ parameters.}
\label{Fig-2}
\end{figure*}

\end{widetext}

Analogously to our first case, we depicted the density parameter, as well as the WEC, DEC, and SEC energy conditions, whose features can be appreciated in Fig. \ref{Fig-2}. There we observe a slow decreasing of $\rho$ in respect of parameters $n$, and $b$ corresponding to an expansion regime smoother than our first case, such a behavior corroborates with features analyzed by Xu et al. \cite{Yixin/2019}. Moreover,  Eqs.\eqref{36}, \eqref{37}, \eqref{38}, and \eqref{39}, unveil that NEC and DEC are satisfied, while WEC is partially obeyed ($\rho >0$) if $  n \leq -1$, and $ b > -4\,\pi\,$. Yet in the energy conditions, we can also see that SEC is again violated, confirming that our Universe experiences an accelerated phase.
If we consider $b=0$ the model reduces the $f(Q,T)$ gravity to $f(Q)$ which presents a strong non-minimal coupling problem. S. Mandal \cite{Mandal/2020} addressed this model, explaining how different energy conditions with $n>0$ give rise to accelerated expansion due to SEC violations.
 Furthermore, the WEC violation along with positive density, makes this $f(Q,T)$ gravity naturally behaves like scalar-tensor gravity models \cite{Whinnett/2004, Mandal/2020}. Despite the viability of such a theory in respect to the energy conditions, the constrained values of parameter $n$ are out of the phenomenological bounds established in \cite{Simran/2020}. There, Arora et al. used Hubble parameter and Supernovae data sets to constraint $n$ and $b$ as $n\,\in \, (1,4)$ and $b \, \in (0,2)$. Therefore, the constraints here imposed for $n$ and $b$ create a tension in use such a $f(Q,T)$ model as a proper description of gravity. We are going to present some extra comments concerning this tension in the next section.

\section{Comparison with $\Lambda$CDM model}\label{sec5}

As a matter of completeness, let us compare our constraints with the $\Lambda$CDM model. This model is so far the most well succeed to describe the evolution of the Universe at different phases.  A direct way to link an $f(Q,T)$ gravity with the $\Lambda$CDM model consists to take the special case $f(Q,T)=f_{\Lambda}(Q) =-Q$ \cite{Lazkoz/2019}. Such a regime yields to the following energy conditions
\begin{itemize}
\item{NEC: $2 \,(1+q)\,H^{2} \geq 0$} ,
\item{WEC: $3 \,H^{2} \geq 0\,,$ and $2 \,(1+q)\,H^{2} \geq 0$} ,
\item{SEC: $6 \,q\, H^{2}\geq 0$} ,
\item{DEC: $2 \,(2-q)\,H^{2}\geq 0$} .
\end{itemize}
One can observe that all energy conditions are satisfied with the present values of $H$ and $q$ except SEC, corroborating with the description of an accelerated expansion. This behavior is compatible with the first model here analyzed for the $f(Q,T)$ gravity. 

Another interesting cosmological parameter which is bounded by experiments is the equation of state parameter $\omega$. Recent observations from Planck Collaboration inform that  $\omega \simeq -1$ \cite{Planck/2018}. Therefore, the EoS parameter is considered a suitable candidate for comparing our models with $\Lambda$CDM. The EoS parameter ($\omega$) is defined as $\omega = \frac{p}{\rho}$. \\
By taking our previous relations for density, and pressure, we are able to find that the EoS parameter for the model A i.e. $f (Q, T )=m Q + b T$ and model B i.e. $f(Q,T)= Q^{n+1}+ b T$ are respectively written as

\begin{equation} \label{40}
\omega= \frac{4 H_{0}^2 m (3 b q_{0}+8 \pi  (2 q_{0}-1))}{b \left(\frac{6 (3 b+16 \pi ) H_{0}^2 m}{b}+4 H_{0}^2 m (q_{0}+1)-6 H_{0}^2 m\right) }.
\end{equation}
and 
\begin{equation} \label{41}
\omega= \frac{3 b (n q_{0}+n+q_{0})+8 \pi  (2 n (q_{0}+1)+2 q_{0}-1)}{b (n q_{0}+n+q_{0}+4)+24 \pi }.
\end{equation}

\begin{figure}[h!]
\centering
\includegraphics[width=7.5 cm]{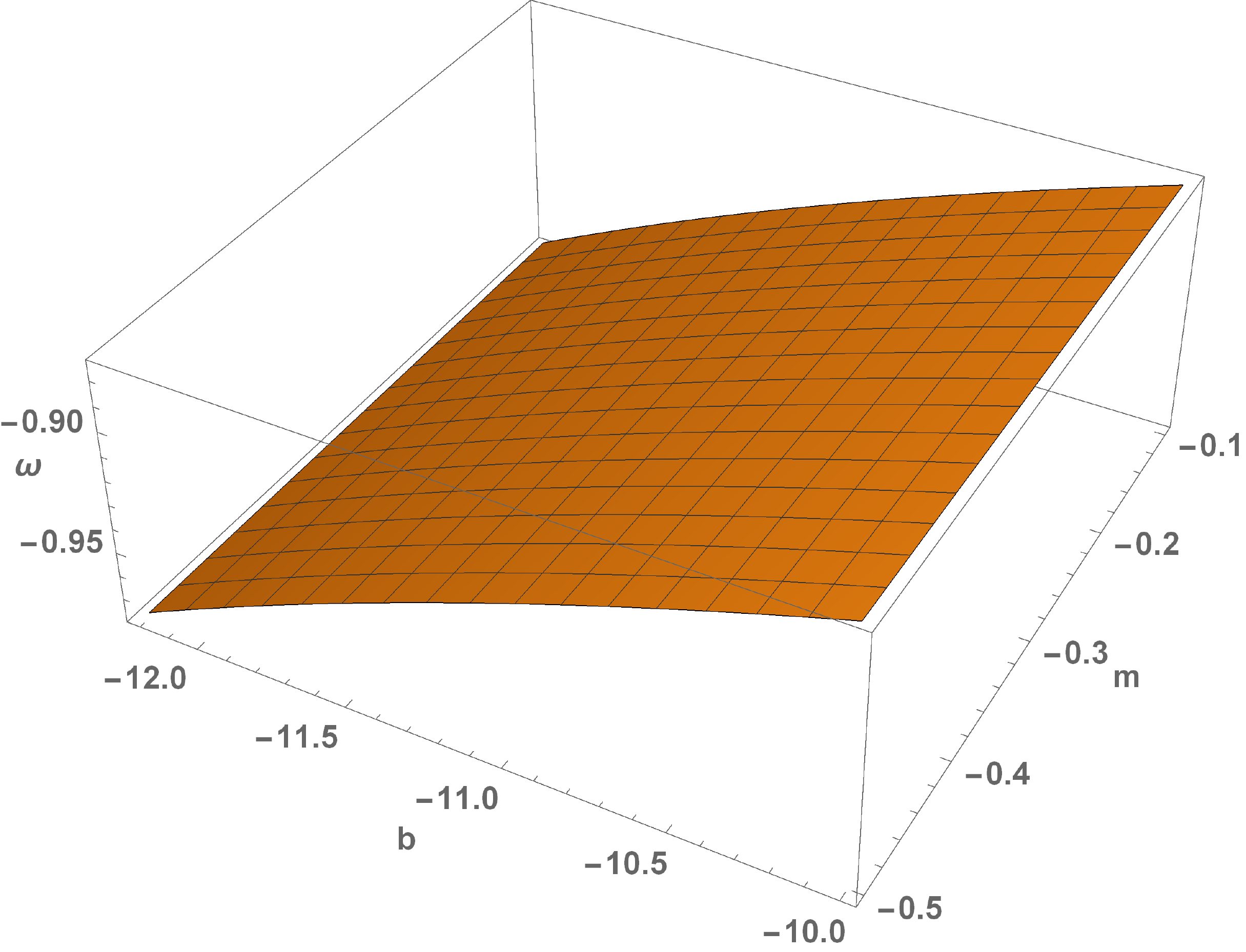}
\caption{EoS parameter for model $f (Q, T )=m Q + b T$ }\label{Fig-OmegaA}
\end{figure}

\begin{figure}[h!]
\centering
\includegraphics[width=7.5 cm]{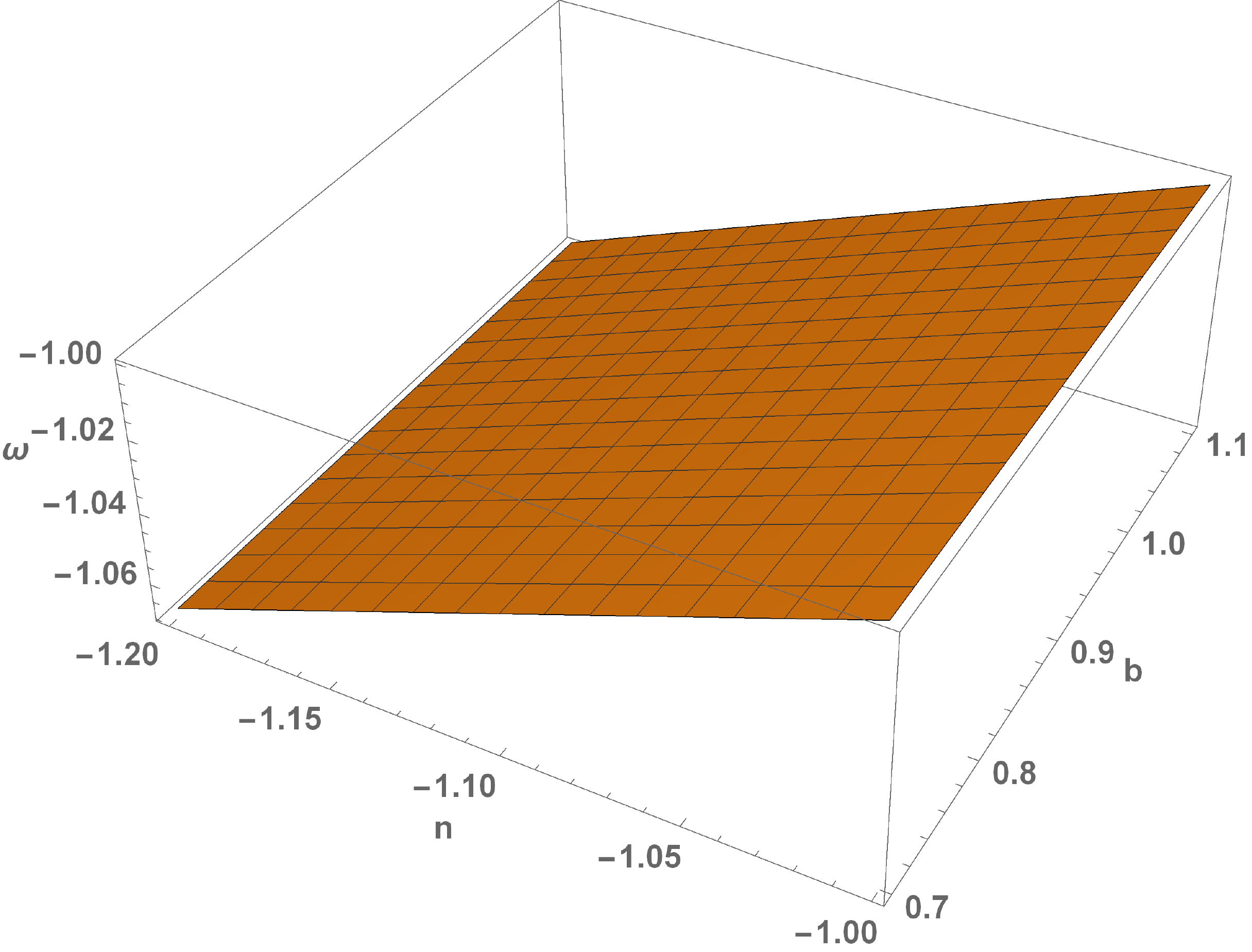}
\caption{EoS parameter for $f(Q,T)= Q^{n+1}+ b T$}\label{Fig-OmegaB}
\end{figure} 

The features of the EoS parameters here derived are presented in Figs \ref{Fig-OmegaA}, and \ref{Fig-OmegaB}. In Fig. \ref{Fig-OmegaA} we have $\omega\approx -1$ from upper values, showing a negative pressure phase compatible with the description of $\Lambda$CDM model. Moreover, in Fig. \ref{Fig-OmegaB} we observe that $\omega \approx -1$ from lower values, surprisingly unveiling a behavior compatible with a phantom era for the dark energy. 

As it is known, the $\Lambda$CDM model, where $\Lambda$ represents a strictly constant vacuum energy, yielding to a negative pressure regime for the Universe, forbids the existence of a phantom era \cite{Cardenas/2020, Carroll/2004, Carroll/2003}. Moreover, a phantom era for the dark energy could lead to a troublesome description for the Universe when we deal with a standard description of gravity plus a fundamental phantom field since DEC energy condition would be violated \cite{Carroll/2004,Sahni/2000}. {Furthermore, in Einstein's gravity the equation of state parameter is constrained to be $ \geq -1$, which means that DEC is satisfied \cite{Carroll/2003}}. However, once the current experimental bounds for $\omega$ establish that
\begin{equation}
\omega = -1.03_{-0.03}^{+0.03}\,,\qquad \mbox {SNe data \cite{Planck/2018}}\,,
\end{equation} 
we still have room for a phantom description of dark energy \cite{Wang/2019}. In the phantom regime the energy density gradually increases, making the Universe accelerates so fast breaking the particles interactions. Overall, there are three categories of phantom regime based on the time evolution of the Hubble parameter, namely \cite{Wang/2019, Frampton/2012}: Big rip if $H(t)\rightarrow \infty$ as $t_{rip}$ is constant - in this scenario even the spacetime rips apart; Little rip if $H(t) \rightarrow \infty$ as $t_{rip}\rightarrow infty$ -  this type of rip prevent singularities in future values of time; Pseudo rip $H(t) \rightarrow \mbox{constant}$ as $t_{rip} \rightarrow \infty$ - such a rip dissociates bound structures that are held together by a binding force. By analyzing the cosmological model for $f(Q,T)= Q^{n+1}+ b T$ studied by Yixin et al. \cite{Yixin/2019}, we realize that
\begin{equation}
H(t) \rightarrow 0\,\qquad \mbox{for} \qquad t_{rip}\rightarrow \infty\,,
\end{equation}
characterizing a compatible pseudo rip evolution for the constrained values of parameters $b$, and $n$ here presented. 

Withal, despite these issues concerning the phantom related with $\Lambda$CDM model, here we show that the$f(Q,T)= Q^{n+1}+ b T$ gravity satisfies DEC if $  n \leq -1$, and $ b > -4\,\pi\,$. Therefore, such a theory of gravity naturally enables us to describe a phantom era for the dark energy, without the need of extra dimensions or phantom scalar fields. The phantom phase is also suggested to reconcile the tension between local and global measurements of the current Hubble value $H_{0}$\cite{Amr/2019}. Hence, we believe that the tension between this $f(Q,T)$ model and observational data for the Hubble parameter at different redshifts, lies in the compatibility of this $f(Q,T)$ with a phantom era description for the dark energy. 

\section{Conclusion}\label{sec6}

An essential role to establish a consistent theory of gravity is the energy condition. As new theories of gravity are bubbling in the literature, it is relevant to put them up to test through constraints over different energy conditions. In this work, we computed the strong, the weak, the null, and the dominant energy conditions for two $f(Q,T)$ gravity models. The $f(Q,T)$ is a promising new theory for gravity based on the combination of the non-metricity function $Q$ with the trace of the energy-momentum $T$.

The models here considered were proposed by  Xu et al. \cite{Yixin/2019}, and constrained by observational data of the Hubble parameter in \cite{Simran/2020}. Firstly we worked with $f(Q,T) = m Q + b T $, where $m$, and $b$ are free parameters. The energy conditions yield us to constraint these free parameters as $b > - 4 \,\pi$, and $m \leq 0$. The previous values result in the violation of SEC, corroborating with an accelerated phase of expansion for the Universe. Besides, such a model is suitable to describe the Universe with respect to energy conditions as $\Lambda$CDM. 

As a second case, we worked with $Q^{n+1}+ b T$, whose free parameters should be constrained to $n \leq -1$, and $ b > -4\,\pi\,$, to satisfy DEC, and NEC energy conditions. In this case, WEC energy condition is partially obeyed while SEC is again violated. The violation of WEC makes this model naturally behaves like scalar-tensor gravity theories. Moreover, the model is compatible with the dark energy era once SEC is not satisfied. A surprisingly feature comes from the equation of state parameter for this model, which describes a phantom regime for the dark energy, allowing extra acceleration for the expansion of the Universe without violates DEC. 

The results here presented allowed us to verify the viability of different families of $f(Q,T)$ gravity models, lighting new paths for a complete description of gravity compatible with the dark energy era, which embeds effects from the quantum era of the Universe. The constraints for our free parameters yield to several testable families for $f(Q,T)$ gravity, opening space even for models compatible with a phantom regime for the dark energy. Moreover, it would be interesting to investigate carefully the coupling of $f(Q,T)$ with inflation fields or with dust, looking for possible analytic models or for cosmological parameters constraints. It would be also interesting to impose constraints on such theories of gravity with observational data from low redshifts, such as BAO measurements at $z=0.1 - 2.5$ which are expected to be performed in the near feature by BINGO \cite{Bingo/2019}, and CHIME \cite{Zhang/2019} telescopes. We hope to report on some of these investigations in the near future.

\acknowledgments S. A. acknowledges CSIR, Govt. of India, New Delhi, for awarding Junior Research Fellowship. JRLS would like to thank CNPq (Grant no. 420479/2018-0), CAPES, and PRONEX/CNPq/FAPESQ-PB (Grant nos. 165/2018, and 0015/2019) for financial support. PKS acknowledges CSIR, New Delhi, India for financial support to carry out the Research project[No. 03(1454)/19/EMR-II Dt.02/08/2019].



\begin{thebibliography}{90}

\bibitem{riess_98} A. G. Riess et al., The Astron. J., {\bf 116} (1998) 1009.

\bibitem{perl_99} G. Perlmutter et al., ApJ, {\bf 517}  (1999) 565.

\bibitem{adler_95} Ronald J. Adler, Brendan Casey, and Ovid C. Jacob, Am. J. Phys, {\bf 63}  (1995) 620.

\bibitem{Planck/2018} N. Aghanim et al. [Planck Collaboration], Planck 2018 results. VI. Cosmological
parameters, astro-ph.CO/1807.06209.

\bibitem{Des/2018} E. Baxter et al. [Dark Energy Survey], Dark Energy Survey Year 1 Results, astroph.CO/1802.05257.

\bibitem{lv_papers} B. P. Abbott et al. (LIGO Scientific Collaboration and Virgo Collaboration),  Phys. Rev. Lett., {\bf 116} (2016) 061102; B. P. Abbott et al. (LIGO Scientific Collaboration and Virgo Collaboration), Phys. Rev. Lett.,  {\bf 119} (2017) 161101; B. P. Abbott et al. (LIGO Scientific Collaboration and Virgo Collaboration), Phys. Rev. Lett.,  {\bf 123} (2019) 011102.

\bibitem{eht_papers} The Event Horizon Telescope Collaboration et al.,  ApJL,  {\bf 875} (2019) L1; The Event Horizon Telescope Collaboration et al.,  ApJL,  {\bf 875} (2019) L5. 

\bibitem{Jimenez/2018} J. B. Jim\'enez, L. Heisenberg, T. Koivisto, Phys. Rev. D, \textbf{98} (2018) 044048.

\bibitem{Lazkoz/2019} R. Lazkoz et al.,  Phys. Rev. D, {\bf 100} (2019) 104027. 

\bibitem{Harko_Lobo/2018} T. Harko, F.S.N. Lobo, Extensions of $f(R)$ gravity: Curvature- Matter Couplings and Hybrid Metric-Palatini Theory (Cambridge University Press, Cambridge 2018).

\bibitem{Harko/2011} T. Harko et al., Phys. Rev. D, {\bf 84} (2011) 024020.

\bibitem{T/2018} T. Harko et al., Phys. Rev. D, {\bf 98} (2018) 084043.

\bibitem{Tharko/2019} T. Harko et al.,  arXiv:1901.00805.

\bibitem{Delhom/2020} A. Delhom, Eur. Phys. J. C, {\bf 80} (2020) 728.

\bibitem{Thomas/2009} Thomas P Sotiriou, Class. Quantum Grav., {\bf 26} (2009) 152001.

\bibitem{Mandal/2020} Sanjay Mandal, P.K. Sahoo, and J.R.L. Santos, Phys. Rev. D, {\bf 102} (2020)  024057.

\bibitem{Yixin/2019} Y. Xu et al.,  Eur. Phys. J. C,  \textbf{79}  (2019) 708.

\bibitem{Simran/2020} Simran Arora, et al., Physics of the Dark Universe, {\bf 30} (2020) 100664.

\bibitem{Sahoo/epjc_2020} Snehasish Bhattacharjee, P. K. Sahoo, Eur. Phys. J. C,  {\bf 80} (2020) 289.

\bibitem{Capozziello/2018} S. Capozziello et al.,  Phys. Lett. B, {\bf 781}  (2018) 99.

\bibitem{Jose} J. B. Jim\'{e}nez et al.,  Phys. Rev. D, {\bf 101}  (2020) 103507.

\bibitem{Kaczmarek/2020}Kaczmarek, A.Z., Szczesniak, Sci Rep, {\bf 10} (2020) 18076.

\bibitem{Yixin/2020} Yixin Xu, Tiberiu Harko, Shahab Shahidi, Shi-Dong Liang, Eur. Phys. J. C, {\bf 80} (2020) 449.

\bibitem{Yang/2021} Jin-Zhao Yang, Shahab Shahidi, Tiberiu Harko, Shi-Dong Liang, arXiv:2101.09956 [gr-qc].

\bibitem{Damianos/2018} Damianos Iosifidis, Christos G. Tsagas, Anastasios C. Petkou, Phys. Rev. D, {\bf 98} (2018) 104037.


\bibitem{Santos} J. Santos et al.,  Phys. Rev. D, \textbf{76}  (2007) 083513.

\bibitem{Rubakov/2014} V. A. Rubakov, Physics-Uspekhi, {\bf 57} (2014) 128. 

\bibitem{Tipler/1978} Frank J. Tipler, Phys. Rev. D, {\bf 17} (1978) 25218.

\bibitem{Crislane/2017} Crislane S. Santos et al., Gen. Relativ. Gravit., {\bf 49} (2017) 50.
 
\bibitem{Sharif} M. Sharif et al.,  Eur. Phys. J. C, \textbf{76}  (2016) 640.

\bibitem{Capozziello/2019} S. Capozziello et al., Int. J. Mod. Phys. D, {\bf 28} (2019) 1930016.

\bibitem{Visser/2000} M. Visser, C. Barcelo,  COSMO-99, (2000) 98.

\bibitem{Whinnett/2004} A. W. Whinnett, D. F. Torres,  The Astrophys. J, {\bf 603} (2004) L133.

\bibitem{Cardenas/2020} V\'ictor H. C\'ardenas, et al., Phys. Rev. D, {\bf 101}, 083530 (2020).

\bibitem{Carroll/2004} Sean M. Carroll, Antonio De Felice, and Mark Trodden, Phys.Rev.D, {\bf 71} (2005) 023525.

\bibitem{Carroll/2003} S. M. Carroll, M. Hoffman, and M. Trodden, Phys. Rev. D, {\bf 68} (2003) 023509.

\bibitem{Sahni/2000} V. Sahni, A.A. Starobinsky, Int. J. Mod. Phys. D, {\bf 9} (2000) 373.

\bibitem{Amr/2019} Amr El-Zant et al., The Astrophys. J, {\bf 871} (2019) 210.

\bibitem{Wang/2019} Jun-Chao Wang, and Xin-He Meng, Eur. Phys. J. C, {\bf 79} (2019) 848.

\bibitem{Frampton/2012}P. H. Frampton, K. J. Ludwick, R. J. Scherrer, Phys. Rev. D, {\bf 85}  (2012) 083001.

\bibitem{Bingo/2019} M. W. Peel, et al. J. Astron. Instrum., {\bf 08} (2019) 1940005.

\bibitem{Zhang/2019} Jing-Fei Zhang, et al. Phys. Lett B, {\bf 799} (2019) 135064.


\end{thebibliography}
\end{document}